\documentclass[11pt,a4paper]{article}
\usepackage{jheppub}
\usepackage[T1]{fontenc}
\usepackage{amssymb}
\usepackage{mathtools}
\usepackage{stackengine}
\usepackage{scalerel}

\usepackage{hyperref}
\usepackage{epsfig}
\usepackage{amsmath,latexsym,amssymb}
\usepackage{graphicx}
\usepackage{braket}
\usepackage{pdfsync}
\usepackage{framed}
\usepackage{mathtools}

\usepackage{jheppub}
\usepackage[T1]{fontenc}
\usepackage{amssymb}
\usepackage{mathtools}
\usepackage{amsmath}
\usepackage{bm}
\usepackage{stackengine}

\usepackage{scalerel}
\usepackage{eufrak}
\usepackage{framed}

\usepackage{mathtools}

\usepackage{pstricks}

\usepackage{soul}

\usepackage{amsmath}
\usepackage{bbold}
\usepackage{bm}
\usepackage{latexsym}
\usepackage{braket}
\usepackage{slashed}
\usepackage{graphicx,booktabs,multirow}
\usepackage{eufrak}
\numberwithin{equation}{section}
\usepackage{color}
\usepackage{pstricks}
\usepackage{amssymb}

\makeatletter
\newcommand{\doublewidetilde}[1]{{%
  \mathpalette\double@widetilde{#1}%
}}
\newcommand{\double@widetilde}[2]{%
  \sbox\z@{$\m@th#1\widetilde{#2}$}%
  \ht\z@=.9\ht\z@
  \widetilde{\box\z@}%
}
\makeatother

\usepackage{hyperref}
\usepackage{slashed}
\usepackage{epsfig}
\usepackage{amsmath,latexsym,amssymb}
\usepackage{graphicx}
\usepackage[latin1]{inputenc}
\usepackage{braket}
\usepackage{pdfsync}
\usepackage{framed}

\usepackage{mathtools}

\def\be{\begin{equation}}
\def\ee{\end{equation}}
\def\ba{\begin{eqnarray}}
\def\ea{\end{eqnarray}}

\def\IC{{\bf C}}

\def\IC{{\bf C}}

\newcommand{\comment}[1]{}

\def\fc#1#2{{\frac{#1}{#2}}}
\newcommand{\req}[1]{(\ref{#1})}

\newcommand{\eea}{\end{eqnarray}}
\def\lf{\left}
\def\ri{\right}

\def\ra{{\rightarrow}}

\author{
Stephan Stieberger${}^{1,2}$, Tomasz R.\ Taylor${}^{3,4}$,\, Bin Zhu${}^{5}$\\[0.5cm]
$^1${\it Max--Planck--Institut f\"{u}r Physik,	Werner--Heisenberg--Institut, 85748 Garching, Germany}\\
$^2${\it Kavli Institute for Theoretical Physics, Santa Barbara, CA 93106, USA}\\
 $^3${\it Department of Physics,
  Northeastern University, Boston, MA 02115, USA}\\
  $^4${\it Faculty of Physics, University of Warsaw, ul. Pasteura 5, 02-093 Warsaw, Poland}\\
$^5${\it School of Mathematics and Maxwell Institute for Mathematical Sciences,\\ University of Edinburgh,
EH9 3FD, UK }\\[0.2cm]
}

\emailAdd{stephan.stieberger@mpp.mpg.de}
\emailAdd{taylor@neu.edu}
\emailAdd{bzhu@exseed.ed.ac.uk}
\title{Carrollian Amplitudes from Strings}

\abstract{ Carrollian holography  is supposed to describe gravity in four--dimensional asymptotically flat space--time by the three--dimensional Carrollian CFT living at null infinity.
We transform superstring scattering amplitudes into the correlation functions  of primary  fields of Carrollian CFT depending on the three--dimensional  coordinates of the celestial sphere and a retarded time coordinate. The power series in the inverse string tension is converted to  a whole tower of 
both UV and IR finite descendants of the underlying field--theoretical Carrollian amplitude.
We focus on four-point amplitudes involving gauge bosons and gravitons in type I open superstring theory and in closed heterotic superstring theory at the tree-level. We also discuss the limit of infinite retarded time coordinates, where the string world--sheet becomes celestial.}

\gdef\@fpheader{}
\makeatother

\begin{document}
\maketitle

\section{Introduction}

Flat space--time holography has gained much interest in recent years.
There are mainly two approaches in the study of flat space--time holography.
The most developed one is celestial holography, which aims to formulate quantum gravity in four--dimensional asymptotically flat space--time in terms of a putative conformal field theory living on the celestial sphere at null infinity.
See \cite{Strominger:2017zoo,Raclariu:2021zjz,Pasterski:2021rjz,Pasterski:2021raf,Donnay:2023mrd} for reviews of celestial holography.
The second approach is Carrollian holography, where a putative three-dimensional Carrollian CFT lives at null infinity $\mathcal{I}$. 
Carrollian CFTs are field theories with BMS symmetries. They can be obtained from standard relativistic CFTs by taking the speed of light to zero, i.e. $c \rightarrow 0$ \cite{Leblond}.
%Only until quite recently, many properties of Carrollian CFTs have been revealed due their connections with flat holography. See for example \cite{}.

Celestial amplitudes \cite{Pasterski:2017kqt,Pasterski:2017ylz}, obtained from standard scattering amplitudes, have been the central objects in celestial holography. 
Scattering amplitudes formulated w.r.t. the  standard momentum eigenstate basis are converted into the boost eigenstate basis making conformal properties manifest. 
As for Carrollian holography, 
it has been shown in  \cite{Donnay:2022aba,Bagchi:2022emh, Donnay:2022wvx} that one can recast scattering amplitudes of massless particles into the so--called Carrollian amplitudes, written in terms of asymptotic or null data at $\mathcal{I}$. 
%See \cite{} for some recent studies on Carrollian amplitudes.
Carrollian amplitudes can be interpreted as correlation functions of Carrollian operators at null infinity.
They are related to the modified  celestial amplitudes \cite{Banerjee:2018gce,Banerjee:2019prz,Banerjee:2020kaa} that appeared in the study of celestial amplitudes.
Perhaps the most inspiring result from these studies is that these two approaches to flat holography are related to each other.
One would hope that some of the difficulties in one approach might be easier to solve  by the other approach. 

Compared to celestial holography where many properties of celestial amplitudes have been understood, we still lack a good understanding of Carrollian amplitudes even though the inspiring paper of Mason, Ruzziconi, and Srikant \cite{Mason:2023mti} has pushed it much forward.
Only a limited amount of examples of Carrollian amplitudes have been computed \cite{Donnay:2022aba,Bagchi:2022emh, Donnay:2022wvx,Salzer:2023jqv,Nguyen:2023miw,Mason:2023mti,Liu:2024nfc,Have:2024dff}.
One of the missing examples are Carrollian string amplitudes, which are the main focus of this work in addition to mixed amplitudes involving gauge bosons and gravitons
 (Einstein--Yang--Mills amplitudes).

In \cite{Stieberger:2018edy}, two of us computed celestial four-point, tree-level amplitudes of gauge bosons and gravitons in type I open superstring theory and in closed heterotic superstring theory.
Several interesting properties of the celestial string amplitudes were observed,
including a factorized $\alpha'$ dependence,
how the single-valued projection that relates heterotic and open string amplitudes is implemented in celestial string amplitudes,
how to take the field theory limit from the celestial string amplitudes,
and a limit where the string world--sheet is identified as the celestial sphere. 

In this work, adapting the prescription shown in \cite{Donnay:2022aba,Bagchi:2022emh, Donnay:2022wvx}, we compute Carrollian four-point, tree level amplitudes from string amplitudes. 
Similar to \cite{Stieberger:2018edy}, we focus on amplitudes of gauge bosons and gravitons in type I open superstring theory and in closed heterotic superstring theory.
In particular, we check whether those interesting properties of the celestial string amplitudes mentioned above still exist in the corresponding Carrollian amplitudes.
We hope to provide some new insights to Carrollian holography, similar to what \cite{Stieberger:2018edy} offered to celestial holography.

\section{Carrollian amplitudes: preliminary}
In \cite{Donnay:2022aba,Bagchi:2022emh, Donnay:2022wvx,Nguyen:2023miw,Mason:2023mti}, the authors showed that scattering amplitudes of massless particles can be streamlined into 3D CFT correlators by Fourier transforms. 
These 3D CFT correlators are called Carrollian amplitudes, served as the CFT data of the putative Carrollian CFT living at null infinity $\mathcal{I} \simeq \mathbf{R}\times S^2$  with coordinates $(u,z,\bar z)$. Here  $(z,\bar z)$ are coordinates on the celestial sphere  $S^2$ and $u$ is a null or retarded time coordinate. 

The first step towards Carrollian amplitudes is to parametrize the light-like momenta by
\be
p^{\mu} = \omega q^{\mu} =  \frac{1}{2}\, \omega(1+|z|^2, z+\bar{z}, -i (z-\bar{z}), 1-|z|^2) \, ,
\ee
where  $\omega$ is the light-cone energy. The amplitudes are expressed in terms of spinor products
\be
\langle ij\rangle= \sqrt{\omega_i\omega_j}\; z_{ij} \ \ ,\ \  \quad [ij] = -\sqrt{\omega_i\omega_j}\;\bar{z}_{ij}
\ee
and the usual scalar products
\be
s_{ij} = 2 p_ip_j = \omega_i\omega_j z_{ij}\bar{z}_{ij} \, .
\ee
The Carrollian amplitudes are obtained by performing Fourier transforms  with respect to the energies\footnote{Notice that the energies are bounded from 0, rather than $-\infty$. More precisely, these transforms are Laplace transforms.  They are related to the modified Mellin transform \cite{Banerjee:2018gce} by setting the conformal dimension $\Delta$ to $1$ \cite{Nguyen:2023miw}. }:
\begin{equation}
\begin{split}
        C_n(\{u_1, z_1,\bar{z}_1\}^{\epsilon_1}_{J_1}, \dots,&\{u_n, z_n,\bar{z}_n\}^{\epsilon_n}_{J_n} ) \\
        &= \prod_{i=1}^n \left( \int_0^{+\infty} \frac{d\omega_i}{2\pi} e^{i\epsilon_i \omega_i u_i}\right)\mathcal{A}_n(\{\omega_1, z_1,\bar{z}_1\}^{\epsilon_1}_{J_1},\dots,\{\omega_n, z_n,\bar{z}_n\}^{\epsilon_n}_{J_n}) \, .  \label{eq:nCarrollian}
\end{split}
\end{equation}
where $\epsilon=\pm 1$ corresponds to outgoing $(+1)$ or incoming $(-1)$ particle and $J$ denotes the particle helicities.
$\mathcal{A}_n$ are scattering amplitudes in momentum basis.  

As explained in \cite{Nguyen:2023miw,Mason:2023mti}, it is also  necessary to consider the descendants by taking derivatives with respect to $u$ coordinates. The $\partial_u-$ descendants of conformal Carrollian primaries are also primaries. Adopting the notations of \cite{Mason:2023mti}, we have
\begin{align}
&C_n^{m_1\dots m_n}(\{u_1, z_1,\bar{z}_1\}^{\epsilon_1}_{J_1}, \dots,\{u_n, z_n,\bar{z}_n\}^{\epsilon_n}_{J_n}) = \partial_{u_1}^{m_1}\dots \partial_{u_n}^{m_n}C_n(\{u_1, z_1,\bar{z}_1\}^{\epsilon_1}_{J_1}, \dots,\{u_n, z_n,\bar{z}_n\}^{\epsilon_n}_{J_n}) \nonumber\\
&= \prod_{i=1}^n \left( \int_0^{+\infty} \frac{d\omega_i}{2\pi} \, (i\epsilon_i \omega_i)^{m_i} e^{i\epsilon_i \omega_i u_i}\right)\mathcal{A}_n(\{\omega_1, z_1,\bar{z}_1\}^{\epsilon_1}_{J_1},\dots,\{\omega_n, z_n,\bar{z}_n\}^{\epsilon_n}_{J_n}) \, ,
\end{align}
and a shorthand notation for $C_n^{1\dots1}$:
\be
\widetilde{C}_n(\{u_1, z_1,\bar{z}_1\}^{\epsilon_1}_{J_1}, \dots,\{u_n, z_n,\bar{z}_n\}^{\epsilon_n}_{J_n}) = C_n^{1\dots1}(\{u_1, z_1,\bar{z}_1\}^{\epsilon_1}_{J_1}, \dots,\{u_n, z_n,\bar{z}_n\}^{\epsilon_n}_{J_n}) \, . \label{eq:tildeC_n}
\ee
As a warm-up, we review the calculations of two-point Carrollian amplitudes shown in \cite{Donnay:2022wvx,Liu:2022mne}. The 2-point tree-level scattering amplitudes in momentum basis are given by
\be
\mathcal{A}_2(\{\omega_1, z_1,\bar{z}_1\}^{-}_{J_1},\{\omega_2, z_2,\bar{z}_2\}^{+}_{J_2}) = \kappa^2_{J_1,J_2}\pi \frac{\delta(\omega_1-\omega_2)}{\omega_1}\delta^{(2)}(z_1-z_2) \delta_{J_1,J_2} \, ,
\ee
where $\kappa_{J_1,J_2}$ is the normalization that depends on the particles involved. Using Eq.(\ref{eq:nCarrollian}), the corresponding two-point Carrollian amplitudes are simply
\begin{align}
&C_2(\{u_1, z_1,\bar{z}_1\}^{-}_{J_1}, \{u_2, z_2,\bar{z}_2\}^{+}_{J_2}) \nonumber\\
&= \frac{1}{4\pi^2} \int_0^{+\infty} d\omega_1 \int_0^{+\infty} d\omega_2 e^{-i\omega_1 u_1} e^{i\omega_2u_2} \mathcal{A}_2(\{\omega_1, z_1,\bar{z}_1\}^{-}_{J_1},\{\omega_2, z_2,\bar{z}_2\}^{+}_{J_2}) \nonumber\\
&= \frac{\kappa^2_{J_1,J_2}}{4\pi} \int_0^{+\infty} \frac{d\omega}{\omega} e^{-i\omega(u_1-u_2)} \delta^{(2)}(z_1-z_2)\delta_{J_1,J_2} \, .
\end{align}
The integral in the last line 
\be
\mathcal{I}_0(u_1-u_2) = \int_0^{+\infty} \frac{d\omega}{\omega} e^{-i\omega(u_1-u_2)} \label{eq:I0u12}
\ee
is divergent and needs to be regulated as \cite{Donnay:2022wvx,Liu:2022mne}
\be
\mathcal{I}_\beta(x) = \lim_{\epsilon\rightarrow 0^+} \int_0^{+\infty} d\omega\, \omega^{\beta-1} e^{-i\omega x-\omega \epsilon} = \lim_{\epsilon\rightarrow 0^+} \frac{\Gamma(\beta)(-i)^{\beta}}{(x-i\epsilon)^{\beta}} \, . \label{eq:Mellin_E_to_omegax}
\ee
In the limit $\beta\rightarrow 0^+$, one finds
\be
\mathcal{I}_{\beta} = \frac{1}{\beta}-\left(\gamma_E+\ln|u_{12}| +\frac{i\pi}{2}\text{sign}(u_{12})\right) +\mathcal{O}(\beta) \, ,
\ee
where $\gamma_E$ is the Euler-Mascheroni constant. As a result, the two-point Carrollian amplitudes take the following form
\begin{multline}
    C_2(\{u_1, z_1,\bar{z}_1\}^{-}_{J_1}, \{u_2, z_2,\bar{z}_2\}^{+}_{J_2}) \\
    = \lim_{\beta\rightarrow0}\frac{\kappa^2_{J_1,J_2}}{4\pi}\left[ \frac{1}{\beta}-\left(\gamma_E+\ln|u_{12}| +\frac{i\pi}{2}\text{sign}(u_{12})\right)\right]\delta^{(2)}(z_{12})\delta_{J_1,J_2} \, .
\end{multline}
 Note the singularity at $\beta \rightarrow 0$ which, according to \cite{Nguyen:2023miw}, is related to the $\ln(r)$ anomaly in the asymptotic expansion of the bulk two-point function. 

From Eq.(\ref{eq:tildeC_n}), the first $\partial_u$ descendant has a simple form
\be
\widetilde{C}_2(\{u_1, z_1,\bar{z}_1\}^{-}_{J_1}, \{u_2, z_2,\bar{z}_2\}^{+}_{J_2}) = \lim_{\epsilon\rightarrow 0^+} \frac{\kappa^2_{J_1,J_2}}{4\pi} \frac{1}{(u_{12}-i\epsilon)^2}\delta^{(2)}(z_{12})\delta_{J_1,J_2} \, .
\ee
Notice that the divergent part $1/\beta$ in $C_2$ is not present in $\widetilde{C}_2$.
%which is divergent and contains $\delta$-function in $z_{12}$ and $\bar{z}_{12}$.

\section{Carrollian amplitudes from field theory}
All three-point Carrollian amplitudes have been computed in \cite{Mason:2023mti}. Moreover, the stringy corrections to the amplitudes are absent for three points due to kinematic reasons.  In this section, we will start from four--point amplitudes. For later use, we will display the four---point Carrollian amplitudes in Yang-Mills (YM) and Einstein-Yang-Mills (EYM) theory.
\subsection{Four-gluon amplitudes}
In the case of four particles, the momentum-conserving delta functions can be written as 
\begin{align}
\delta^{(4)}(\omega_1 q_1 +\omega_2 q_2 &-\omega_3 q_3-\omega_4 q_4) =  \frac{4}{\omega_4 |z_{14}|^2 |z_{23}|^2} \nonumber\\
&\times \delta\left( \omega_1 -\frac{z_{24} \bar{z}_{34}}{z_{12} \bar{z}_{13}} \omega_4 \right) \delta\left( \omega_2 -\frac{z_{14} \bar{z}_{34}}{z_{12}\bar{z}_{32}} \omega_4\right) \delta\left(\omega_3 + \frac{z_{24}\bar{z}_{14}}{z_{23}\bar{z}_{13}}\omega_4\right) \delta(r-\bar{r}) \, , \label{eq:4deltas}
\end{align}
where $r$ is the two-dimensional conformal invariant cross ratio:
\be
r = \frac{z_{12}z_{34}}{z_{23}z_{41}} \, .
\ee
Note that we have chosen a specific in and out configuration: particle 1 and 2 are incoming while particle 3 and 4 are outgoing.  The reality constraint on the cross ratio imposed by $\delta(r-\bar{r})$ follows from the momentum conservation law \cite{Pasterski:2017ylz,Stieberger:2018edy}. The physical meaning of the cross ratio can be understood from the relation between Mandelstam's variables $s=s_{12} = (p_1+p_2)^2$, $u=-s_{23} = (p_2-p_3)^2$, and the scattering angle $\theta$:
\be
\frac{s_{23}}{s_{12}} = \frac{1}{r} = -\frac{u}{s} = \sin^2\left( \frac{\theta}{2}\right) \, .
\ee
In the physical domain, $r>1$, $ s>0$, $u=-s/r<0$.

We begin with the well-known four-point MHV gluon amplitude
\be
\mathcal{M}(-,-,+,+) = \frac{\langle 12 \rangle^3}{\langle 23\rangle \langle 34 \rangle \langle 41\rangle}  =  \frac{\omega_1\omega_2}{\omega_3\omega_4} \frac{z_{12}^3}{z_{23}z_{34}z_{41}} = r \, \frac{z_{12}\bar{z}_{34}}{\bar{z}_{12}z_{34}} \, , \label{eq:MHVgluon}
\ee
where in the last step we have used the kinematic constraints from Eq.(\ref{eq:4deltas}). The corresponding Carrollian amplitude is 
\begin{align}
&C_{4,\text{YM}}(-,-,+,+) \nonumber\\
&= \int_0^{+\infty} \frac{d\omega_1}{2\pi} \int_0^{+\infty} \frac{d\omega_2}{2\pi}\int_0^{+\infty} \frac{d\omega_3}{2\pi}\int_0^{+\infty} \frac{d\omega_4}{2\pi} e^{-i\omega_1 u_1}e^{-i\omega_2 u_2}e^{i\omega_3 u_3}e^{i\omega_4 u_4} \, r \, \frac{z_{12}\bar{z}_{34}}{\bar{z}_{12}z_{34}} \nonumber\\
&\times\frac{4}{\omega_4 |z_{14}|^2 |z_{23}|^2}\delta\left( \omega_1 -\frac{z_{24} \bar{z}_{34}}{z_{12} \bar{z}_{13}} \omega_4 \right) \delta\left( \omega_2 -\frac{z_{14} \bar{z}_{34}}{z_{12}\bar{z}_{32}} \omega_4\right) \delta\left(\omega_3 + \frac{z_{24}\bar{z}_{14}}{z_{23}\bar{z}_{13}}\omega_4\right) \delta(r-\bar{r}) \nonumber\\
&= \frac{1}{4\pi^4} \, \frac{z_{12}\bar{z}_{34}}{\bar{z}_{12}z_{34}} \frac{r\delta(r-\bar{r})}{|z_{14}|^2 |z_{23}|^2} \int_0^{+\infty} \frac{d\omega_4}{\omega_4} e^{i\omega_4 x_4} \, , \label{eq:C4YM}
\end{align}
where we have defined
\begin{align}
x_4 &=  u_4- \frac{z_{24} \bar{z}_{34}}{z_{12} \bar{z}_{13}} u_1-\frac{z_{14} \bar{z}_{34}}{z_{12}\bar{z}_{32}} u_2-\frac{z_{24}\bar{z}_{14}}{z_{23}\bar{z}_{13}} u_3  \nonumber\\
&= u_4 - \frac{r}{r-1} \frac{|z_{24}|^2}{|z_{12}|^2} u_1 -\frac{1}{r} \frac{|z_{34}|^2}{|z_{23}|^2} u_2 + (r-1) \frac{|z_{14}|^2}{|z_{13}|^2}u_3 \, . \label{eq:defx4}
\end{align}
The integral in the last line of Eq.(\ref{eq:C4YM}) is the same as the one in Eq.(\ref{eq:I0u12}), which needs to be regulated as explained there. Therefore, 
\be
C_{4,\text{YM}}(-,-,+,+) =\frac{1}{4\pi^4} \, \frac{z_{12}\bar{z}_{34}}{\bar{z}_{12}z_{34}} \frac{r\delta(r-\bar{r})}{|z_{14}|^2 |z_{23}|^2} \, \mathcal{I}_0(-x_4) \, .
\ee
Similar to the two-point case, we compute  the $u$--descendant $\widetilde{C}_4$ and we find
\begin{align}
\widetilde{C}_{4,\text{YM}}(-,-,+,+) &=  \partial_{u_1}\partial_{u_2}\partial_{u_3} \partial_{u_4} C_{4,\text{YM}} \nonumber\\
&=-\frac{4}{(2\pi)^4} \frac{z_{12}\bar{z}_{34}}{\bar{z}_{12}z_{34}}\frac{r\delta(r-\bar{r})}{|z_{14}|^2 |z_{23}|^2} \frac{z_{24} \bar{z}_{34}}{z_{12} \bar{z}_{13}}\frac{z_{14} \bar{z}_{34}}{z_{12}\bar{z}_{32}} \frac{z_{24}\bar{z}_{14}}{z_{23}\bar{z}_{13}} \int_0^{+\infty} d\omega_4 \, e^{i\omega_4 x_4} \omega_4^3 \nonumber\\
&=\frac{1}{4\pi^4} \frac{z_{24}^2 \, \bar{z}_{34}^3\, r \,\delta(r-\bar{r})}{|z_{12}|^2 \,z_{34} \, \bar{z}_{13}^2 |z_{23}|^4} \,\left(u_4- \frac{z_{24} \bar{z}_{34}}{z_{12} \bar{z}_{13}} u_1-\frac{z_{14} \bar{z}_{34}}{z_{12}\bar{z}_{32}} u_2-\frac{z_{24}\bar{z}_{14}}{z_{23}\bar{z}_{13}} u_3\right)^{-4}  \nonumber\\
&= \frac{1}{4\pi^4} \frac{z_{24}^2 \, \bar{z}_{34}^3\, r \,\delta(r-\bar{r})}{|z_{12}|^2 \,z_{34} \, \bar{z}_{13}^2 |z_{23}|^4} \,\, x_4^{-4} \, .
\end{align}
Our results agree with \cite{Mason:2023mti} up to an overall constant.

\subsection{Four-point mixed gauge-gravitational amplitudes}
The simplest amplitude with one graviton and three gluons in EYM  is \cite{Stieberger:2016lng}
\be
\mathcal{M}(--,-,+,+) = \frac{\langle 12\rangle^4}{\langle 23 \rangle \langle 34 \rangle \langle 42\rangle} = \frac{\omega_1^2\omega_2}{\omega_3\omega_4} \frac{z_{12}^4}{z_{23}z_{34}z_{42}} = r \frac{z_{12} \bar{z}_{34}^2 z_{14}}{\bar{z}_{12} z_{34} \bar{z}_{13}} \omega_4 \, .
\ee
Proceeding the same way as before, we obtain the corresponding Carrollian amplitude
\be
C_{4,\text{EYM}}(--,-,+,+) = \frac{i}{4\pi^4} \frac{z_{12} \bar{z}_{34}^2 r\delta(r-\bar{r})}{\bar{z}_{12}z_{34} \bar{z}_{13}\bar{z}_{14} |z_{23}|^2} \frac{1}{x_4} \, , \label{eq:C4EYM}
\ee
with $x_4$ defined in Eq. (\ref{eq:defx4}). Note, that this amplitude \req{eq:C4EYM} has not been computed before in the literature and represents the first example of a mixed Carrollian amplitude involving both a graviton and several gauge bosons.

\subsection{Four-graviton amplitudes}
The four-graviton amplitudes can be obtained from the pure Yang-Mill amplitudes by using the famous KLT relations \cite{Kawai:1985xq}. The tree-level 4-graviton amplitude reads
\be
\mathcal{M}(--,--,++,++) = \mathcal{M}(-,-,+,+)\; s_{12}\; \mathcal{M}'(-,-,+,+) \, , \label{eq:4graviton}
\ee
where $\mathcal{M}(-,-,+,+)$ is the Yang-Mills amplitudes given by Eq.(\ref{eq:MHVgluon}) and the prime implies exchanging $3$ and $4$ with respect to the canonical $(1,2,3,4)$ ordering. In terms of the energy and celestial coordinates, this can be written as
\be
\mathcal{M}(--,--,++,++) = \omega_4^2 \frac{|z_{14}|^2 |z_{34}|^2}{|z_{13}|^2} \left( r \frac{z_{12}\bar{z}_{34}}{\bar{z}_{12} z_{34}} \right)^2 \,.
\ee
The corresponding Carrollian amplitude is 
\be
C_{4,\text{E}}(--,--,++,++) =- \frac{\delta(r-\bar{r}) |z_{34}|^2}{|z_{23}|^2 |z_{13}|^2} \left( r \frac{z_{12}\bar{z}_{34}}{\bar{z}_{12} z_{34}} \right)^2 \frac{1}{x_4^2}\ , \label{eq:C4graviton}
\ee
 which is IR finite in contrast to the gluon amplitude Eq.\req{eq:C4YM}.
Moreover, in contrast to the celestial graviton amplitude \cite{Stieberger:2018edy} the Carrollian graviton amplitude is UV finite as the time coordinate $u$ acts as a UV regulator.

\section{Carrollian amplitudes from string theory}
\subsection{Four-gluon amplitudes in open superstring theory}

First, we consider type I open superstring. The type I open superstring amplitude is related to the Yang-Mills amplitude Eq.(\ref{eq:MHVgluon}) by a simple rescaling
\be
\mathcal{M}_I(-,-,+,+) = \mathcal{M}(-,-,+,+) F_I(s,u) \, ,
\ee
with the string ``formfactor'' \cite{Green:1981xx}
\be
F_I(s,u) = -\alpha' s_{12} B(-\alpha' s_{12}, 1+\alpha' s_{23}) = -s B(-s,1-u) = \frac{\Gamma(1-s)\Gamma(1-u)}{\Gamma(1-s-u)}\, , \label{eq:FI}
\ee
where we have rescaled Mandelstam's variables by the string scale: $s=\alpha' s_{12}$ and $u=-\alpha' s_{23} = -s/r$. Upon using the momentum conserving delta function (\ref{eq:4deltas}), we have
\begin{align}
s &= \alpha' (r-1) \frac{|z_{14}|^2 |z_{34}|^2}{|z_{13}|^2} \omega_4^2 \, , \label{eq:Mands} \\
u&= \alpha' \frac{1-r}{r} \frac{|z_{14}|^2 |z_{34}|^2}{|z_{13}|^2} \omega_4^2 \, ,\label{eq:Mandu}\\
s+u &= \alpha' \frac{(r-1)^2}{r} \frac{|z_{14}|^2 |z_{34}|^2}{|z_{13}|^2} \omega_4^2 \, .\label{eq:Mands+u}
\end{align}
In this section,
we will follow three different ways of evaluating the corresponding Carrollian amplitudes.

\subsubsection{Carrollian string amplitude as series over field theory descendants}

In the first method, starting from the fact that the beta function admits a Laurent expansion (cf.  e.g.  \cite{Sondow07,Brown:2019wna}),
 one can rewrite the string formfactor in Eq.(\ref{eq:FI}) as follows
\be
F_I(s,u)=\frac{\Gamma(1-s)\Gamma(1-u)}{\Gamma(1-s-u)}\,= \exp\left\{ \sum_{n\geq 2} \frac{\zeta(n)}{n} \left(s^n+u^n-(s+u)^n\right)\right\} \, , \label{eq:FIexp}
\ee
where $\zeta(n)$ are Riemann zeta values. Expanding the exponential, it gives the $\alpha'$ expansions:
\be
F_I(s,u) = 1-\zeta(2) su - \zeta(3) s\, u(s+u) -\frac{2}{5} \zeta(2)^2 (s^2 +\frac{1}{4} su +u^2) +\mathcal{O}(\alpha'^5) \, .\label{ExpansionBeta}
\ee
We proceed to compute the Carrollian amplitude by using Eq.(\ref{eq:FIexp}):
\begin{align}
&C_{4I}(-,-,+,+) =  \nonumber\\
=& \int_0^{+\infty} \frac{d\omega_1}{2\pi} \int_0^{+\infty} \frac{d\omega_2}{2\pi}\int_0^{+\infty} \frac{d\omega_3}{2\pi}\int_0^{+\infty} \frac{d\omega_4}{2\pi} \;e^{-i\omega_1 u_1}e^{-i\omega_2 u_2}e^{i\omega_3 u_3}e^{i\omega_4 u_4} \, r \, \frac{z_{12}\bar{z}_{34}}{\bar{z}_{12}z_{34}} \nonumber\\
&\times\frac{4}{\omega_4 |z_{14}|^2 |z_{23}|^2}\delta\left( \omega_1 -\frac{z_{24} \bar{z}_{34}}{z_{12} \bar{z}_{13}} \omega_4 \right) \delta\left( \omega_2 -\frac{z_{14} \bar{z}_{34}}{z_{12}\bar{z}_{32}} \omega_4\right) \delta\left(\omega_3 + \frac{z_{24}\bar{z}_{14}}{z_{23}\bar{z}_{13}}\omega_4\right) \delta(r-\bar{r}) F_I(s,u)\nonumber\\
=&\frac{1}{4\pi^4}\frac{z_{12}\bar{z}_{34}}{\bar{z}_{12}z_{34}} \frac{r\delta(r-\bar{r})}{|z_{14}|^2 |z_{23}|^2} \int_0^{+\infty} d\omega_4\; e^{i\omega_4 x_4} \frac{1}{\omega_4} \exp\left\{ \sum_{n\geq 2} \frac{\zeta(n)}{n} \left(s^n+u^n-(s+u)^n\right)\right\} \, , \label{eq:C4I1}
\end{align}
where $s$, $u$, and $s+u$ are given by Eqs.(\ref{eq:Mands}) to (\ref{eq:Mands+u}). Notice that $s$, $u$, and $s+u$ are all proportional to $\omega_4^2$. Since the integrand contains an exponential $e^{i\omega_4 u_4}$, we can trade each $\omega_4^2$ for a derivative $\partial_{u_4}^2$ and pull the exponential out of the integral, leading to
\begin{align}
&C_{4I}(-,-,+,+) \nonumber\\
=& \exp\left\{ \sum_{n\geq2} \frac{\zeta(n)}{n} \frac{\alpha'^{\,n} |z_{14}|^{2n} |z_{34}|^{2n}}{|z_{13}|^{2n}} \left[(r-1)^n+\left(\frac{1-r}{r}\right)^n-\left(\frac{(r-1)^2}{r}\right)^n \right] (-1)^n\partial_{u_4}^{2n}\right\} \nonumber\\
&\times \frac{1}{4\pi^4} \, \frac{z_{12}\bar{z}_{34}}{\bar{z}_{12}z_{34}} \frac{r\delta(r-\bar{r})}{|z_{14}|^2 |z_{23}|^2} \int_0^{+\infty} \frac{d\omega_4}{\omega_4} e^{i\omega_4 x_4} \, , \nonumber\\
=&\exp\left\{ \sum_{n\geq2} \frac{\zeta(n)}{n} \frac{\alpha'^{\,n} |z_{14}|^{2n} |z_{34}|^{2n}}{|z_{13}|^{2n}} \left[(r-1)^n+\left(\frac{1-r}{r}\right)^n-\left(\frac{(r-1)^2}{r}\right)^n \right](-1)^n\partial_{u_4}^{2n}\right\} \nonumber\\
&\times C_{4,\text{YM}}(-,-,+,+) \, , \label{eq:C4Imethod1}
\end{align}
where we used Eq.(\ref{eq:C4YM}) for the Carrollian gluon amplitudes from YM. 
This result shows that the Carrollian amplitudes from type I open string theory can be obtained by a differential operator acting on the field theory Carrollian amplitudes.
The differential operator $\partial_{u_4}$ generates the $\partial_u$-descendants of the field theory Carrollian amplitudes thus encoding the information of the $\alpha'$--expansions.
Notice that although the $\partial_{u_4}$ derivatives  in (\ref{eq:C4Imethod1}) increase conformal dimensions, the whole combination in the exponent of (\ref{eq:C4Imethod1}) does not change the conformal dimension of $C_{4,\text{YM}}$ due to  the compensating conformal weights supplied by the $|z_{ij}|$ factors. 
Note, that this way a whole  tower of  both UV and IR finite  descendants of the field theory Carrollian amplitude is generated.

Comparing Eq. (\ref{eq:C4Imethod1}) with its celestial counterpart in \cite{Stieberger:2018edy} shows, that the $\alpha'$ dependence of (\ref{eq:C4Imethod1}) does not factorize.

\subsubsection{Carrollian string amplitude as series over Nielsen polylogarithms}

Next, we  shall present the second method of computing the Carrollian amplitude in type I open string. As we will see, it can be written as a double series expansion.
It is convenient to use the integral representation of the string formfactor Eq.(\ref{eq:FI}) 
\be
F_I(s,u) =-sB(-s,1-u) = -s\int_0^1 dx\, x^{-1-s}(1-x)^{-u} \, .
\ee
The Carrollian amplitudes Eq.(\ref{eq:C4I1}) becomes
\begin{align}
&C_{4I}(-,-,+,+) \nonumber\\
=&\frac{\alpha' (1-r)}{4\pi^4}\frac{|z_{34}|^2}{|z_{13}|^2}\frac{z_{12}\bar{z}_{34}}{\bar{z}_{12}z_{34}} \frac{r\delta(r-\bar{r})}{ |z_{23}|^2}\int_0^1\frac{dx}{x} \int_0^{+\infty} d\omega_4 \, e^{i\omega_4 x_4} \omega_4\, \exp\left[-\rho\, \omega_4^2 [r\ln x-\ln (1-x)]\right] \, , \label{eq:C4Imethod2a}
\end{align}
where
\be
\rho = \frac{\alpha' (r-1)}{r} \frac{|z_{14}|^2 |z_{34}|^2}{|z_{13}|^2} \, . \label{eq:rho}
\ee
The integral in Eq.(\ref{eq:C4Imethod2a}) can be written as 
\begin{align}
&\int_0^1\frac{dx}{x}\int_0^{+\infty} d\omega_4 \, e^{i\omega_4 x_4} \omega_4\, \exp\left[-\rho\, \omega_4^2 [r\ln x-\ln (1-x)]\right]  \nonumber\\
&=\sum_{k=0}^{\infty} \int_0^{+\infty} \,d\omega_4\, \omega_4^{1+2k} \, e^{i\omega_4 x_4} \int_0^1 \frac{dx}{x} \frac{(-1)^k \rho^k}{k!} [r\ln x-\ln (1-x)]^k  \nonumber\\
&= \sum_{k=0}^{\infty} \frac{(-1)^k\, \rho^k\, \Gamma(2+2k)}{(ix_4)^{2+2k}} \sum_{l=0}^k r^{l} \int_0^1 \frac{dx}{x} \frac{(\ln x)^l [- \ln(1-x)]^{k-l}}{l! (k-l)!} \, , \label{eq:C4Idx}
\end{align}
where we used Eq.(\ref{eq:Mellin_E_to_omegax}) and binomial expansion $(a-b)^k = \sum_{l=0}^k {k \choose l}  a^l (-b)^{k-l}$ in the last line. The  integral w.r.t. $x$ in Eq.(\ref{eq:C4Idx}) is related to Nielsen's polylogarithm functions $S_{n,k}(t)$ \cite{Kolbig:1983qt}
\be
S_{n,k}(t) = \frac{(-1)^{n+k-1}}{(n-1)! k!}\int_0^1 \frac{dx}{x} \ln^{n-1}x\; \ln^k(1-xt) \, , \quad t\in \mathbf{C} \, ,
\ee
labeled by positive integers $n$ and $k$. To simplify the notation, we denote $S(n,k) = S_{n,k}(1)$. The series expansion in (\ref{eq:C4Idx}) can be written as 
\begin{align}
&\sum_{k=1}^{\infty} \frac{(-1)^k\, \rho^k\, \Gamma(2+2k)}{(ix_4)^{2+2k}} \sum_{l=0}^{k-1} (-r)^{l} S(l+1,k-l)  \nonumber\\
&+ \sum_{k=0}^{\infty}\frac{(-1)^k\, \rho^k\, \Gamma(2+2k)}{(ix_4)^{2+2k}} r^k \int_0^{1} \frac{dx}{x} \frac{\ln^k x}{k!} \, , \label{eq:C4Idxseries}
\end{align}
where the second line corresponds to the $l=k$ terms in Eq.(\ref{eq:C4Idx}). As we will see, the second line of Eq.(\ref{eq:C4Idxseries}) reproduces the field--theory Carrollian amplitude Eq.(\ref{eq:C4YM}):
\begin{align}
&\sum_{k=0}^\infty\frac{(-1)^k\, \rho^k\, \Gamma(2+2k)}{(ix_4)^{2+2k}} r^k \int_0^{1} \frac{dx}{x} \frac{\ln^k x}{k!}  \nonumber\\
=& \sum_{k=0}^\infty  \int_0^{+\infty} d\omega_4\, e^{i\omega_4 x_4} \, \omega_4^{1+2k} \int_0^{1} \frac{dx}{x} \frac{(-\rho r \ln x)^k}{k!} \nonumber\\
=& \int_0^{+\infty} d\omega_4 e^{i\omega_4 x_4} \omega_4 \int_0^1 \frac{dx}{x} e^{-\rho r \omega_4^2 \ln x} = -\int_0^{+\infty} d\omega_4 e^{i\omega_4 x_4} \frac{1}{\rho r \omega_4}  \, .
\end{align}
Combined with the prefactor in Eq.(\ref{eq:C4Imethod2a}) and use Eq.(\ref{eq:rho}), this part reproduces exactly the field--theory Carrollian amplitude from Eq.(\ref{eq:C4YM}).
Therefore, the Carrollian amplitudes in type I open string theory, computed by using this method, is given by
\begin{align}
C_{4I}&(-,-,+,+) = C_{4,\text{YM}}(-,-,+,+)\, \, \nonumber\\
&+\frac{\alpha' (1-r)}{4\pi^4}\frac{|z_{34}|^2}{|z_{13}|^2}\frac{z_{12}\bar{z}_{34}}{\bar{z}_{12}z_{34}} \frac{r\delta(r-\bar{r})}{ |z_{23}|^2}\sum_{k=1}^{\infty} \frac{(-\rho)^k\, \Gamma(2+2k)}{(ix_4)^{2+2k}} \sum_{l=0}^{k-1} (-r)^{l} S(l+1,k-l) \, . \label{eq:C4Imethod2result}
\end{align}
Moreover, the coefficients of the double series expansion are related to zeta values and multiple zeta values in the following way \cite{KolbigZZA,Adamchik}
\begin{align}
S(k,1) &= Li_{k+1}(1) = \zeta(k+1) \,\quad l=k-1 , \\
S(l+1, k-l) &= \zeta(l+2, \{1\}^{k-l-1}) \, , \quad  l<k-1 \, , \label{NPMZV}
\end{align}
where 
\be
\zeta(n+1, \{1\}^{k-1}) = \zeta(n+1,\underbrace{1,\dots, 1}_{k-1}) = \sum_{n_1>n_2>\dots>n_k} \frac{1}{n_1^{n+1} n_2 \cdots n_k} 
\ee
is a multiple zeta value (MZV) of depth $k$.

Notice that since $\rho$ defined in Eq.(\ref{eq:rho}) contains $\alpha'$, meaning all the $\alpha'$ corrections in Eq.(\ref{eq:C4Imethod2result}) to the Carrollian field theory amplitudes are contained in the double series expansion.

\subsubsection{Carrollian string amplitude and Drinfeld associator}
 
 In this part we present a third method, which also relates the other two methods presented above.
Rather than using Eq.\req{eq:FIexp} for the expansion Eq.\req{ExpansionBeta} we use the series \cite{Borwein:1996yq}
\be
F_I(s,u)=1-\sum_{k,l\geq0}c_{kl}\ s^{k+1}u^{l+1}\ ,\label{ExpansionBeta1}
\ee
with 
\be\label{Dcoeffs}
c_{kl}=c_{lk}=\zeta(l+2,\{1\}^k)=\zeta(k+2,\{1\}^l)\ ,\ k,l\geq0\ ,
\ee
being the coefficients of the logarithm of the Drinfeld associator $\Phi(A,B)$. More precisely \cite{Drinfeld}
\be\label{lnDri}
\ln \Phi(A,B)=\sum_{k,l\geq 0} c_{kl}\; {\rm ad}(B)^l{\rm ad}(A)^k[A,B]\;h^{k+l+2}
\ee
modulo second commutants  $[[L,L],[L,L]]$  of the Lie algebra $L$ generated by $A$ and $B$.
Here $h$ is an order parameter and ${\rm ad}(X) \ldots=[X,\ldots]$.
With using the expansion Eq.\req{ExpansionBeta1} we may repeat the step leading to Eq.\req{eq:C4I1} and obtain
\be
C_{4I}(-,-,+,+) =\frac{1}{4\pi^4}\frac{z_{12}\bar{z}_{34}}{\bar{z}_{12}z_{34}} \frac{r\delta(r-\bar{r})}{|z_{14}|^2 |z_{23}|^2} \int_0^{+\infty} \fc{d\omega_4}{\omega_4}\; e^{i\omega_4 x_4} \; \lf\{1-\sum_{k,l\geq0}c_{kl}\ x^{k+1}y^{l+1}\ri\}\ ,\label{leadingto}
\ee
with $x=r\rho\omega_4^2$ and $y=-\rho\omega_4^2$ and $\rho$ defined in Eq.\req{eq:rho}. Clearly, the first term of Eq.\req{leadingto} 
gives rise to the field theory part Eq.\req{eq:C4YM} of the amplitude.
After separating  the latter and performing the $\omega_4$ integration we arrive at
\begin{align}
C_{4I}(-,-,+,+)&=C_{4,\text{YM}}(-,-,+,+) \label{KITP}\\
&+\frac{1}{4\pi^4}\frac{z_{12}\bar{z}_{34}}{\bar{z}_{12}z_{34}} \frac{r\delta(r-\bar{r})}{|z_{14}|^2 |z_{23}|^2}\ \sum_{k,l\geq0}(-1)^k\; c_{kl}\; \fc{\Gamma(4+2k+2l)}{x_4^{4+2k+2l}}\; r^{k+1}\rho^{k+l+2}\ ,\nonumber
\end{align}
which can be cast into:
\begin{align}
C_{4I}(-,-,+,+)&=C_{4,\text{YM}}(-,-,+,+) \label{resultC}\\
&-\frac{1}{4\pi^4}\frac{z_{12}\bar{z}_{34}}{\bar{z}_{12}z_{34}} \frac{r\delta(r-\bar{r})}{|z_{14}|^2 |z_{23}|^2}\ \sum_{k\geq 1}\sum_{l=0}^{k-1} c_{k-l-1,l}\; \fc{\Gamma(2+2k)}{x_4^{2+2k}}\; (-r)^{k-l}\rho^{k+1}\ .\nonumber
\end{align}
The result Eq.\req{resultC} for the Carrollian string amplitude agrees with Eq.\req{eq:C4Imethod2result} subject to Eqs.\req{NPMZV} and \req{Dcoeffs}.
The expression Eq.\req{resultC} represents a double series comprising the coefficients Eq.\req{Dcoeffs} of the (logarithmic) Drinfeld associator Eq.\req{lnDri}. For $r=-1$ the sum over $l$ can be performed explicitly \cite{Sondow07}:
$$\sum_{l=0}^{k-1}\zeta(k-l+1,\{1\}^l)=\fc{I_{k-1}}{(k-1)!}\ ,$$
with the integral
$$I_n=\int_\Delta dxdy\;\fc{(-\ln xy)^n}{xy}$$
over the triangle $\Delta=\{(x,y)\}\in[0,1]^2\;|\;x+y\geq1\}$, e.g. $I_0=\zeta(2),\; I_1=2\zeta(3)$.

Note, that the Drinfeld associator $\Phi(A,B)$, which captures  the monodromy of a universal version of the Knizhnik--Zamolodchikov (KZ) equation, appears in knot theory and in conformal field theory \cite{Knots}.  In the latter the KZ equation describes a Ward identity of CFT correlators of primaries \cite{DiFrancesco:1997nk}. Therefore, it would be interesting to derive \req{KITP} directly in the underlying Carrollian CFT from a differential equation w.r.t.\ Carrollian coordinates $(u,z,\bar z)$ resembling a KZ equation.

\subsection{Four-gluon amplitudes in heterotic superstring theory}

In heterotic superstring theory, similarly to type I, the four-gluon amplitude is related to the Yang-Mills amplitude Eq.(\ref{eq:MHVgluon}) by a simple rescaling,
\be
\mathcal{M}_{H}(-,-,+,+) = \mathcal{M}(-,-,+,+)\ F_H(s,u) \, ,
\ee
where the heterotic formfactor is \cite{Gross:1985rr}
\be
F_H(s,u) = -\frac{\Gamma(-\alpha' s_{12})\Gamma(\alpha' s_{23}) \Gamma(\alpha's_{31})}{\Gamma(\alpha' s_{12})\Gamma(-\alpha's_{23})\Gamma(-\alpha' s_{31})} =- \frac{\Gamma(-s)\Gamma(-t)\Gamma(-u)}{\Gamma(s)\Gamma(t)\Gamma(u)} \, . \label{eq:FH}
\ee
Similar to Eq.(\ref{eq:FIexp}), the heterotic formfactor can be written as
\be
F_H(s,u) =  \exp\left\{ \sum_{{n\geq2 \atop n \, \text{odd}}} \frac{2\zeta(n)}{n} \left(s^n+u^n-(s+u)^n\right)\right\} \, , \label{eq:FHexp}
\ee
In \cite{Stieberger:2013wea}, it has been observed that Eq.(\ref{eq:FHexp}) can be obtained from Eq.(\ref{eq:FIexp}) by applying the following single-valued projection:
\begin{align}
\text{sv}:
\begin{cases}
			\zeta(2n+1) \rightarrow 2\zeta(2n+1), & \text{ $n\geq1$ }\\
            \zeta(2n) \rightarrow 0. & \text{}
		 \end{cases}
\end{align}
The relation between open and closed string amplitudes through the  single-valued projection has been established in \cite{Stieberger:2014hba}.
The four--point Carrollian gluon amplitude in this case is written as the following integral
\be
C_{4H}(-,-,+,+) = \frac{1}{4\pi^4}\frac{z_{12}\bar{z}_{34}}{\bar{z}_{12}z_{34}} \frac{r\delta(r-\bar{r})}{|z_{14}|^2 |z_{23}|^2} \int_0^{+\infty} d\omega_4\; e^{i\omega_4 x_4} \frac{1}{\omega_4} F_H(s,u) \, , \label{eq:C4H}
\ee
with $x_4$, $s$ , and $u$ given by Eqs.(\ref{eq:defx4}), (\ref{eq:Mands}) and (\ref{eq:Mandu}).
Following the steps of the first method in the previous subsection, we obtain  the Carrollian amplitude 
\begin{align}
C_{4H}&(-,-,+,+)=\nonumber \\
&= \exp\left\{ \sum_{{n\geq2 \atop n \, \text{odd}}} \frac{2\zeta(n)}{n} \frac{\alpha'^{\,n} |z_{14}|^{2n} |z_{34}|^{2n}}{|z_{13}|^{2n}} \left[(r-1)^n+\left(\frac{1-r}{r}\right)^n-\left(\frac{(r-1)^2}{r}\right)^n \right](-1)^n\partial_{u_4}^{2n}\right\} \\  \nonumber
&\times C_{4,\text{YM}}(-,-,+,+)\label{eq:C4Hmethod1}\end{align}
which can be obtained from Eq.(\ref{eq:C4Imethod1}) by the single-valued projection.

Next, we shall present the second way of computing the Carrollian  gluon amplitude in heterotic string theory. The string formfactor Eq.(\ref{eq:FH}) can be written as a complex integral:
\be
F_H(s,u) = -\frac{s}{\pi} \int d^2 z |z|^{-2s-2} |1-z|^{-2u} (1-z)^{-1} \, .
\ee
Substituting Eqs.(\ref{eq:Mands}) and Eqs.(\ref{eq:Mandu}),  Eq.(\ref{eq:C4H}) becomes
\begin{align}
C_{4H}(-,-,+,+) =&\frac{\alpha' (1-r)}{4\pi^4}\frac{|z_{34}|^2}{|z_{13}|^2}\frac{z_{12}\bar{z}_{34}}{\bar{z}_{12}z_{34}} \frac{r\delta(r-\bar{r})}{ |z_{23}|^2}  \nonumber\\ 
&\times \frac{1}{\pi} \int \frac{d^2 z}{|z|^2 (1-z)} \int_0^{+\infty} d\omega_4 \, e^{i\omega_4 x_4} \,\omega_4  \, \exp\left[-\omega_4^2\left[ \rho r \ln |z|^2 -\rho \ln |1-z|^2\right] \right]\, , \label{eq:CHmethod2}
\end{align}
where $\rho$ is defined in Eq.(\ref{eq:rho}). After expanding the exponential w.r.t. $\omega_4^2$ and following similar steps than those leading to  Eq. (\ref{eq:C4Idx}) the integral becomes: 
\begin{align}
& \frac{1}{\pi} \int \frac{d^2 z}{|z|^2 (1-z)} \int_0^{+\infty} d\omega_4 \, e^{i\omega_4 x_4} \,\omega_4  \, \exp\left[-\omega_4^2\left[ \rho r \ln |z|^2 -\rho \ln |1-z|^2\right] \right] \nonumber\\
=& \sum_{k=0} \frac{1}{\pi} \int \frac{d^2 z}{|z|^2 (1-z)} \int_0^{+\infty} d\omega_4 e^{i\omega_4 x_4} \, \omega_4^{1+2k} \frac{(-1)^k \rho^k}{k!} \left[ r \ln |z|^2 -\ln|1-z|^2\right]^k   \nonumber\\
=& \sum_{k=0}^\infty \frac{(-\rho)^k \Gamma(2+2k)}{(i x_4)^{2+2k}} \sum_{l=0}^k r^l \frac{1}{\pi} \int \frac{d^2z}{|z|^2 (1-z)} \frac{(\ln|z|^2)^l(-\ln|1-z|^2)^{k-l}}{l!(k-l)!} \, . \label{eq:CHd2z}
\end{align}
Once again, as we will see, the $l=k$ terms correspond to the field theory Carrollian amplitude.
The other terms are related to the generalized single-valued Nielsen's polylogrithms:
\be
S^{\bf c}_{n,k}(t)\equiv \fc{(-1)^{n+k-1}}{\pi(n-1)!\,k!}\
\int_\IC \fc{d^2z}{|z|^2}\ (1-z)^{-1}\ \ln^{n-1}|z|^2\ \ln^k
|1-zt|^2\  .
\ee
where $n$ and $k$ are positive integers. We define $S^{\bf c}(n,k):= S^{\bf c}_{n,k}(1)$. Then, Eq.  (\ref{eq:CHd2z}) becomes
\begin{align}
&\sum_{k=1}^{\infty} \frac{(-\rho)^k \Gamma(2+2k)}{(i x_4)^{2+2k}} \sum_{l=0}^{k-1} (-r)^l S^{\bf{c}}(l+1,k-l)  \nonumber\\
&+ \sum_{k=0}^{\infty} \frac{(-\rho)^k \Gamma(2+2k)}{(ix_4)^{2+2k}} \frac{r^k}{\pi k!} \int d^2z \frac{\ln^k|z|^2}{|z|^2(1-z)} \, , \label{eq:CHmethod2stepb}
\end{align}
where the last line contains the $k=l$ terms in Eq.(\ref{eq:CHd2z}). Similar to the type I open string case, we will see that this part reproduces the field theory Carrollian amplitude from YM Eq.(\ref{eq:C4YM}).

In polar coordinate $z=x e^{i\phi}$, the angular integral of the last line in Eq.(\ref{eq:CHmethod2stepb}) becomes
\begin{align}
\frac{1}{\pi}\int_0^1 d\phi\; (1-x e^{i\phi})^{-1} = 
\begin{cases}
	2\, ,		 & 0 <x<1\\
             0 \, , &  x>1
		 \end{cases}
\end{align}
Then, the last line of Eq. (\ref{eq:CHmethod2stepb}) becomes
\begin{align}
&2\sum_{k=0}^{\infty} \frac{ \Gamma(2+2k)}{(ix_4)^{2+2k}}   \int_0^1 \frac{dx}{x} \frac{(-2\rho r \ln x)^k}{k!} =2\sum_{k=0}^{\infty} \int_0^{+\infty} d\omega_4 e^{i\omega_4 x_4} \omega_4^{1+2k} \int_0^1 \frac{dx}{x} \frac{(-2\rho r \ln x)^k}{k!} \nonumber\\
&= 2 \int_0^{+\infty} d\omega_4 \, e^{i\omega_4 x_4} \omega_4 \int_0^1 \frac{dx}{x} e^{-2\omega_4^2 \,\rho \,r \ln x} = -\int_0^{+\infty} d\omega_4 \, e^{i\omega_4 x_4} \frac{1}{\rho\, r\omega_4} \, ,
\end{align}
Plugging it back to Eqs.(\ref{eq:CHmethod2stepb}) and (\ref{eq:CHmethod2}), we obtain
\begin{align}
C_{4H}&(-,-,+,+) = C_{4,\text{YM}}(-,-,+,+)\, \, \nonumber\\
&+\frac{\alpha' (1-r)}{4\pi^4}\frac{|z_{34}|^2}{|z_{13}|^2}\frac{z_{12}\bar{z}_{34}}{\bar{z}_{12}z_{34}} \frac{r\delta(r-\bar{r})}{ |z_{23}|^2}\sum_{k=1}^{\infty} \frac{(-1)^k\, \rho^k\, \Gamma(2+2k)}{(ix_4)^{2+2k}} \sum_{l=0}^{k-1} (-r)^{l} S^{\bf c}(l+1,k-l) \, . \label{eq:C4Hmethod2result}
\end{align}
Moreover, the coefficients of the double series expansion are related to zeta values and multiple zeta values in the following way
\begin{align}
S^{\bf c}(k,1) &= \text{sv}(\zeta(k+1)) \, \quad l=k-1 \\
S^{\bf c}(l+1,k-l) &= \text{sv}(\zeta(l+2,\{1\}^{k-l-1})) \, \quad l< k-1  \, ,
\end{align}
where the single-valued map can be found in \cite{Brown:2013gia}.

%\begin{align}
%\cases{\zeta(2n+1)\mapsto 2\ \zeta(2n+1),& $n\geq1$\ ,\cr
%\zeta(2)\mapsto 0\ .&}
%\end{align}

\subsection{Mixed gauge-gravitational amplitudes in heterotic superstring theory}
The amplitude with one graviton and three gauge bosons also exists in heterotic superstring theory.
Similarly to the case of four gluons, it is related to EYM amplitude by a simple rescaling \cite{Stieberger:2016lng,Schlotterer:2016cxa}
\be
\mathcal{M}_H(--,-,+,+) = \mathcal{M}(--,-,+,+)\ F_H(s,u) \, ,
\ee
where $F_H$ is the same heterotic formfactor Eq.(\ref{eq:FH}) as in the Yang-Mills case.
We proceed to the computation of the corresponding Carrollian amplitude in the same way as in the previous subsection. We find
\begin{align}
&C_{4H}(--,-,+,+) \nonumber\\
=& \exp\left\{ \sum_{{n\geq2 \atop n \, \text{odd}}} \frac{2\zeta(n)}{n} \frac{\alpha'^{\,n} |z_{14}|^{2n} |z_{34}|^{2n}}{|z_{13}|^{2n}} \left[(r-1)^n+\left(\frac{1-r}{r}\right)^n-\left(\frac{(r-1)^2}{r}\right)^n \right](-1)^n\partial_{u_4}^{2n}\right\}  \nonumber\\
&\times C_{4,\text{EYM}}(--,-,+,+) \, , \label{eq:C4Hmixmethod1}
\end{align}
where $C_{4,\text{EYM}}(--,-,+,+)$ is given by Eq.(\ref{eq:C4EYM}). Equivalently,
\begin{align}
C_{4H}&(--,-,+,+)=C_{4,\text{EYM}}(--,-,+,+) \,  \, \nonumber\\
&+ \frac{\alpha' (1-r)}{4\pi^4} \frac{z_{12} z_{14} \bar{z}_{34}^3 r\delta(r-\bar{r})}{\bar{z}_{12}\bar{z}_{13} |z_{13}|^2 |z_{23}|^2}\sum_{k=1}^{\infty} \frac{(-1)^k\, \rho^k\, \Gamma(3+2k)}{(-ix_4)^{3+2k}} \sum_{l=0}^{k-1} (-r)^{l}\; S^{\bf c}(l+1,k-l) \, . \label{eq:C4Hmixmethod2}
\end{align}

\subsection{Graviton amplitudes in heterotic superstring theory}

In heterotic superstring theory, the four--graviton amplitude is related to Einstein's amplitude by a simple rescaling,
\be
\mathcal{M}_{H}(--,--,++,++) = \mathcal{M}(--,--,++,++)\ F_H(s,u) \, ,
\ee
where $F_H(s,u)$ is the same string formfactor Eq.(\ref{eq:FH}) as in the four-gluon case. As a result, we find the corresponding Carrollian amplitude
\begin{align}
&C_{4H}(--,--,++,++) \nonumber\\
=& \exp\left\{ \sum_{{n\geq2 \atop n \, \text{odd}}} \frac{2\zeta(n)}{n} \frac{\alpha'^{\,n} |z_{14}|^{2n} |z_{34}|^{2n}}{|z_{13}|^{2n}} \left[(r-1)^n+\left(\frac{1-r}{r}\right)^n-\left(\frac{(r-1)^2}{r}\right)^n \right](-1)^n\partial_{u_4}^{2n}\right\}  \nonumber\\
&\times C_{4,\text{E}}(--,--,++,++) \, , \label{eq:C4Hgramethod1}
\end{align}
with $C_{4,\text{E}}(--,--,++,++)$ is given in Eq.(\ref{eq:C4graviton}).
And equivalently:
\begin{align}
&C_{4H}(--,--,++,++)=C_{4,\text{E}}(--,--,++,++) \,  \, \nonumber\\
&+ \frac{\alpha' (1-r)}{4\pi^4} \frac{\delta(r-\bar{r}) |z_{14}|^2 |z_{34}|^4}{|z_{23}|^2 |z_{13}|^4} \left( r\frac{z_{12}\bar{z}_{34}}{\bar{z}_{12} z_{34}} \right)^2\sum_{k=1}^{\infty} \frac{(-1)^k\, \rho^k\, \Gamma(4+2k)}{(ix_4)^{4+2k}} \sum_{l=0}^{k-1} (-r)^{l} S^{\bf c}(l+1,k-l) \, . \label{eq:C4Hmixmethod2}
\end{align}

\section{Saddle-point approximation in Carrollian amplitudes from strings}
In this section, we will show that in certain limit of the $u$ coordinate of the Carrollian string amplitudes, the string world sheet becomes the celestial sphere in a similar way to section 4 of \cite{Stieberger:2018edy}.
We adapt the notation used in section 4 of \cite{Stieberger:2018edy} and denote $a=1/r$. In the region of $s<0$ and $u=-as <0$, the string formfactor of type I open string, see Eq.(\ref{eq:FI}),  can be written as
\be
F_I(s,u) =-s B(-s,1-u) =-s \int_0^1 dx\; x^{-1-s}(1-x)^{as} \, .
\ee
In order to discuss the $s\rightarrow -\infty$ limit, we write
\be
-sB(-s,1-u) = -s\int_0^1 dx\; x^{-1} e^{-s f(x)} \, , \quad f(x) = \ln x- a\ln(1-x) \, . \label{eq:FIexpsec5}
\ee
Using the steepest descent (saddle point) method \cite{Gross:1987kza} and solving the saddle point equation, one finds
\be
f'(x_0) = 0 \Rightarrow x_0 = \frac{1}{1-a} \, . \label{eq:worldsheetsaddle}
\ee
The saddle point approximation (Laplace method) states that the integral for a function $f(x)$ is approximated by
\be
\int_a^b g(x) \exp[\alpha' f(x)]\; dx \sim \sqrt{\frac{2\pi}{\alpha' |f''(x_0)|}}\; [g(x_0) +\mathcal{O}(\alpha'^{-1})] \exp[\alpha' f(x_0)] \, . \label{eq:Laplace}
\ee
The string formfactor Eq.(\ref{eq:FIexpsec5}) becomes \cite{Stieberger:2018edy}
\be
F_I(s,u)\sim \sqrt{\frac{2\pi a s}{1-a}} (-a)^{as} (1-a)^{(1-a)s} \, ,
\ee
which is exponentially suppressed at $s\rightarrow -\infty$ with $a<0$. For the case of the physical range of $s>0$, $u<0$, see section 3 and 4 in \cite{Stieberger:2018edy}.

Switching to the Carrollian corresponding amplitude, we start from Eq.(\ref{eq:C4Imethod2a}), where we encountered an integral of the form 
\be
\int_0^{\infty} d\omega\, \omega^{\nu-1} e^{-\beta \omega^2-\gamma \omega}  = (2\beta)^{-\nu/2}\Gamma(\nu)\exp\left( \frac{\gamma^2}{8\beta}\right) D_{-\nu}\left(\frac{\gamma}{\sqrt{2\beta}}\right) \, ,
\ee
where $D_{-\nu}(x)$ is the parabolic cylinder function. For our case $\nu=2$  we can use the following relation between the parabolic cylinder function $D_{-n-1}$ and the complementary error function $\text{erfc}$:
\be
D_{-n-1}(z) = \sqrt{\frac{\pi}{2}} \frac{(-1)^n}{n!} e^{-\frac{1}{4}z^2} \frac{d^n\left( e^{\frac{1}{2} z^2} \text{erfc}(z/\sqrt{2})\right)}{dz^n} \, .
\ee
As a result, we find 
\be
\int_0^{\infty} d\omega\, \omega\, e^{-\beta \omega^2 -\gamma \omega} = \frac{1}{2\beta} -e^{\frac{\gamma^2}{4\beta}} \sqrt{\pi} \gamma \frac{\text{erfc}\left(\frac{\gamma}{2\sqrt{\beta}}\right)}{4\beta^{3/2}} \, . \label{eq:erfc}
\ee
Concretely, we have
\begin{align}
\beta &= \frac{\rho}{a}\left[  \ln x-a \ln(1-x)\right] \, , \\
\gamma & = -i x_4 \, ,
\end{align}
where $\rho$ and $x_4$ are given by Eqs.(\ref{eq:rho}) and (\ref{eq:defx4}). We notice that the first term in Eq.(\ref{eq:erfc}) does not depend on the Carrollian coordinate $u$.
From the  point of view of Carrollian amplitudes, the second term in Eq.(\ref{eq:erfc}) is more interesting. Hence, we will focus on the second term in the following.

In the region of $a<0$, $\beta>0$, we are interested in the limit $x_4 \rightarrow +\infty$. This can be achieved by taking $u_4, u_2\rightarrow +\infty$ or $u_1, u_3\rightarrow -\infty $. It corresponds to going to the corners of null infinity.
The complementary error function admits the following asymptotic behavior:
\be
\text{erfc}\left(\frac{\gamma}{2\sqrt{\beta}}\right) = 2 + e^{-\frac{\gamma^2}{4\beta}} \left(2\sqrt{\frac{\beta}{2}}\frac{1}{\gamma} +\mathcal{O}\left(\frac{1}{\gamma^3}\right)\right) \, .
\ee
Taking the contribution from the leading term, now the $x$-integral in the Carrollian amplitude Eq.(\ref{eq:C4Imethod2a}) becomes
\be
x_4\int_0^1 \frac{dx}{x} \exp\left(\frac{-a \, x_4^2}{4\rho[\ln x-a \ln(1-x)]}\right) \left[ \frac{\rho}{a} \left(\ln x -a \ln(1-x)\right)\right]^{-3/2} \, ,
\ee
where we omit an overall factor. Applying the Laplace method in Eq.(\ref{eq:Laplace}), we find that in the limit $x_4\rightarrow +\infty$, the integral is dominated by the saddle point at
\be
x_0= \frac{1}{1-a} \, , \label{eq:x_0a}
\ee
which is the same as the  saddle point of string world sheet Eq.(\ref{eq:worldsheetsaddle}).
After some algebra, we obtain a compact expression for the $u$-dependent part of the Carrollian amplitude in type I open string:
\begin{align}
C_{4I,u-\text{dependent}}(-,-,+,+) 
&\sim \frac{a\delta(a-\bar{a})}{|z_{14}|^2|z_{23}|^2} \; \sqrt{\frac{2a}{(1-a)[-(1-a)\ln(1-a)-a\ln(-a)]}}\nonumber\\
&\times \exp\left(\frac{a \, x_4^2}{4 \rho[(1-a)\ln(1-a)+a\ln(-a)]}\right) \, .
\end{align}
To conclude, while in celestial holography the string world--sheet becomes celestial 
in the limit of large dimensions, $\sum_i\lambda_i\ra\infty$, in the Carrolian holography this correspondence is reached   by going to the corners of null infinity.

 It would be interesting to find a physical or geometrical meaning of the saddle point solution (\ref{eq:x_0a}) in the context of Carrollian CFT. As for the physical range of parameters $s > 0$, $u < 0$, the saddle point method would still work with a careful treatment of the integral contour known as the Pochhamer contour \cite{Witten:2013pra,Stieberger:2018edy}.

\section{Discussion}
We have transformed superstring scattering amplitudes into correlation functions of primary fields of the putative Carrollian CFT at null infinity, so-called Carrollian amplitudes.
We have focused on tree-level four-point amplitudes involving gauge bosons and gravitons in type I open superstring theory and in closed heterotic superstring theory. 
The Carrollian amplitudes are presented  in two ways.
In the first way, cf.\ e.g.\  Eq.\ (\ref{eq:C4Imethod1}), the string $\alpha'$--expansions are translated into  a whole tower of both UV and IR finite $u$--descendants of the underlying field theory Carrollian amplitudes.
In the second way, cf.\ e.g.\ Eqs.\ (\ref{eq:C4Imethod2result}) and \req{resultC}, the Carrolllian amplitudes are written as double series expansions with the coefficients determined by Nielsen polylogarithms or coefficients of Drinfeld associator, respectively, and their single--valued version.
We have explained how the single--valued projection, which relates heterotic and open string amplitudes, is implemented in Carrollian amplitudes. The properties of Carrollian amplitudes should  be helpful in extracting more information about the underlying CFT at null infinity.

There are many open questions on  Carrollian amplitudes in general. Here, we list a few of them. First, we expect some of the observed properties  in four-point Carrollian string amplitudes to take over to higher--point amplitudes. We hope to address  this question in the future.

Including loop corrections in celestial amplitudes has not been well understood, see however  \cite{Gonzalez:2020tpi,Arkani-Hamed:2020gyp,Magnea:2021fvy,Ball:2021tmb,Costello:2022upu,Donnay:2022hkf,Pasterski:2022djr,Bhardwaj:2022anh,Bittleston:2022jeq,He:2023lvk,Costello:2023hmi,Donnay:2023kvm,Krishna:2023ukw,Banerjee:2023jne,Tao:2023wls,Chattopadhyay:2024kdq} on some progress.
For the Carrollian approach, four-point Carrollian loop amplitudes have been studied in \cite{Liu:2024nfc} in massless $\phi^4$ theory.
It would be interesting to see if there is a general scheme of including loop corrections to Carrollian amplitudes.

In celestial CFT, there is an interesting relation between celestial MHV gluon amplitudes and Liouville theory \cite{Stieberger:2022zyk,Taylor:2023bzj,Stieberger:2023fju}. It would be  fascinating if one could understand this relation from the Carrollian point of view.

It has been suggested before that Carrollian amplitudes are the natural objects obtained from the flat space limit of AdS amplitudes with suitable boundary conditions.
See \cite{Hijano:2019qmi,Li:2021snj,deGioia:2022fcn,deGioia:2023cbd,Bagchi:2023fbj,Duary:2023gqg,Bagchi:2023cen} for recent works on this direction.
An interesting question is what kind of AdS amplitudes give rise to Carrollian string amplitudes in the flat space limit.

%\begin{itemize}
%\color{blue}
%\item Shall we add some outlook on corrections to the tree-level OPEs, possible generalization to Susy case, or string amplitudes?
%\end{itemize}

\section*{Acknowledgements}
BZ would like to thank Tim Adamo, Wei Bu, Gerben Oling, and Piotr Tourkine for helpful discussions on related topics. This research was supported in part by grant NSF PHY-2309135 to the Kavli Institute for Theoretical Physics (KITP).  
TRT is supported by the National Science Foundation
under Grants Number PHY-1913328 and PHY-2209903, by the 
NAWA Grant 
``Celestial Holography of Fundamental Interactions'' and 
by the Simons Collaboration on Celestial Holography.
Any opinions, findings, and conclusions or
recommendations expressed in this material are those of the authors and do not necessarily
reflect the views of the National Science Foundation. BZ is supported by Royal Society.

\end{document}